%% file: mbs.tex
\documentclass[ASNA,twocolumn]{USG} 
\usepackage{anyfontsize} %

\usepackage{colortbl}   
\usepackage{xcolor}     

\input{aux_format}  

\usepackage{steinmetz}
\graphicspath{{./images/}}

\articletype{RESEARCH ARTICLE}%
\subarticletype{Particle Technology and Fluidization}

\received{00}
\revised{00}
\accepted{00}
\journal{International Journal of Adaptive Control and Signal Processing}
\volume{0}
\copyyear{00}
\startpage{1}
\articledoi{00}

\begin{document}
\title{An Algebraic State Observer for a Self-Sensing Active Magnetic Bearing System}
\transtitle{An Algebraic State Observer for a Self-Sensing Active Magnetic Bearing System}
\author[1]{Olga Zarina}[https://orcid.org/0009-0008-8613-2835]
\author[1]{Natalya Martyukhova}
\author[1]{Alexey Bobtsov}[https://orcid.org/0000-0003-1854-6717]
\author[2]{Romeo Ortega}[https://orcid.org/0000-0002-7747-5405]

\titlemark{An Algebraic State Observer for a Self-Sensing Active Magnetic Bearing System}

\address[1]{\orgdiv{Control Systems and Robotics Department, }\orgname{ITMO University, }%
\orgaddress{\state{Saint-Petersburg, }\country{Russia}}}

\address[2]{\orgdiv{Department of Electrical and Electronic Engineering, }\orgname{ITAM, }%
\orgaddress{\state{Mexico City, }\country{Mexico}}}

\corres{Zarina Olga (\email{oakozachek@itmo.ru})}

\fundingInfo{Supported by the Ministry of Science and Higher Education of the Russian Federation (project no. FSER-2025-0002).}


\transkeywords{Nonlinear algebraic observers | flux observers | self-sensing active magnetic bearing systems}

\abstract[ABSTRACT]{The problem of designing a globally stable observer for a self-sensing active magnetic bearing system assuming only measurements of currents and voltages is addressed in this paper. Towards this end, we first design a radically different, high performance, state observer, which is obtained invoking novel techniques. Indeed, our objective is to obtain an {\em algebraic} relation between the unmeasurable part of the state and filtered versions of the systems inputs and outputs, which holds for all times. Then, using this algebraic observer, we propose a {\em robust asymptotic} version of the observer. Simulation results that illustrate the performance of the observer are also presented.}

 \transabstract[transABSTRACT]{The problem of designing a globally stable observer for a self-sensing active magnetic bearing system assuming only measurements of currents and voltages is addressed in this paper. Towards this end, we first design a radically different, high performance, state observer, which is obtained invoking novel techniques. Indeed, our objective is to obtain an {\em algebraic} relation between the unmeasurable part of the state and filtered versions of the systems inputs and outputs, which holds for all times. Then, using this algebraic observer, we propose a {\em robust asymptotic} version of the observer. Simulation results that illustrate the performance of the observer are also presented.}


\maketitle


\section{Introduction}
\label{sec1}

As it is widely recognized, energy-efficiency targets have become increasingly stringent. Hence, active magnetic bearings (AMBs) have emerged as a prevalent and promising technology, demonstrating outstanding performance and potential in multiple industrial rotating machines, such as compressors, turbines, and generators \cite{MASbook}. AMBs offer significant advantages over other bearing technologies because of the omission of lubricants. Moreover, they can operate in extreme environments such as a vacuum and high-temperature conditions. Traditional AMB systems rely on displacement sensors to measure rotor displacement and achieve feedback control. However, in practical applications, these AMB systems equipped have several shortcomings widely discussed in the literature \cite{MASbook,YANetal}. 

The removal of displacement sensors, that used to provide the required information for the control of the AMBs, posed a new challenging problem referred in the applications journal as {\em Displacement Self-sensing Technology} that, as explained in  \cite[Section 1]{YANetal} boils down to ``directly estimating rotor displacements by processing signals within the actuators." Clearly, in the control theoretic language this reduces to the design of {\em state observers}, a problem that has been extensively studied by the control community, see {\em e.g.}, \cite{ASTKARORTbook,BERbook,BERANDAST}. It is widely recognized that self-sensing technology---equivalently, observer design---has become an important research direction in AMBs and has attracted widespread attention.  

The first ``self-sensing technology`` AMBs was proposed by Vischer, a scholar from the Swiss Federal Institute of Technology in
Zurich, in 1988 \cite{VISbook}---see also \cite{VISBLE}. Vischer’s method was based on the Luenberger observer
scheme for the linearized state-space model of the equilibrium point in a single-degree-of-freedom magnetic levitation system. Since the publication of this research the magnetics community has been very active developing ``self-sensing technology" based on a mixture of observer design theory and refinements of the modeling assumptions. For instance, in \cite{MONbook} it was shown that using a linear periodic model linear Luenberger observers achieve very good performance---basically because the periodic model accounts for the switching ripple effects of the PWM power
amplifier. This is also one of the foundational theoretical studies for AMB self-sensing schemes based on demodulation methods---see also the discussion in \cite{MASbook} and \cite{YANetal}.

Attempts have been made in the magnetics community to apply {\em nonlinear} observers, adaptive  filters and sliding mode observers, but as reported in \cite{YANetal}, the experimental results were usually unsatisfactory. The Kalman filter displacement estimator optimized using a particle swarm algorithm proposed by \cite{SUNZHU} was successfully applied in a laboratory setting for a three-pole, five-degree-of-freedom AMB. However, the Kalman filter algorithm for the state transition model of a high-order (15th-order) AMB poses significant {\em computational challenges} for digital controllers. A recent survey of the studies in observer design for AMBs may be found in \cite{GLUetal,PYRetal_ijc20}, while \cite{DUTetal} contains a very detailed, recent overview of the whole field. 

In this paper we follow the novel approach started in \cite{BOBetal_tac} and center our attention in the design of {\em algebraic observers} for the AMB system. The qualifier ``algebraic" meaning that we want to obtain an {\em algebraic} relation between the system state and filtered versions of the systems inputs and outputs, which holds true {\em for all $t \geq 0$}. The algebraic property should be contrasted with the usual procedure of designing a dynamical system whose output (the state estimate) {\em asymptotically} (or in fixed/finite time) converges to the systems state. Following the procedure first articulated in \cite{BOBetal_tac}, to design this observer we take a radically new approach which relies on the application of the {\em Swapping Lemma}  \cite[Lemma 3.6.5]{SASBODbook}, and name the observer as Algebraic Swapping Lemma Observer (ASLO).\footnote{In a long series of publications, including \cite{FLIJOISIR}, it is claimed that an algebraic observer is derived. However, in all these papers the authors appeal to the use of the {\em practically inadmissible} operation of successive signal differentiation.}

Unfortunately, the usual operating conditions of the AMBs are such that very little ``excitation" is imposed to the system---in the context of ASLO this is tantamount to the requirement that the algebraic equations that define the ASLO can be {\em explicitly} derived---an operation that, in this particular case, involves a matrix inversion. To overcome the problem of lack of excitation in the ASLO, we also present an {\em asymptotic} observer, which is directly obtained from the ASLO and does not require the matrix inversion. It is interesting to note the recent report \cite{BOBetal_jpc} where we presented an ASLO solution to the challenging problem of state estimation of batch reactors. It is well-known \cite{RAPDOC} that, asymptotic observers are inadequate for this kind of systems since they operate on a small fixed time and their dynamics are asymptotically {\em unobservable}.      

The remainder of the paper is organized as follows. Section \ref{sec2} gives the problem formulation.  In Section \ref{sec3} we present the main results, namely an ASLO and an asymptotic state observer based on it, called A-ASLO. Some simulation results are given  in  Section \ref{sec4}. The paper is wrapped-up with concluding remarks and future research directions in  Section \ref{sec5}. A well-known key technical result---the Swapping Lemma \cite[Lemma 3.6.5]{SASBODbook}---is given in Appendix \ref{appa} and the proof of the main result is given in Appendix \ref{appb}.\\

\noindent {\bf Notation} For $b \in \rea^n$ we denote the Euclidean norm as $|b|$. ${\bf I}_n$ is the $n \times n$ identity matrix and ${\bf 0}_{n \times s}$ is an $n \times s$ matrix of zeros. The action of {a linear time-invariant (LTI)} filter $\calf(p) \in \rea(p)$ on a signal $w(t)$ is denoted as $\calf[w]$, where $p^n[w]:={d^n w(t)\over dt^n}$. Given a function $H:  \rea^n \to \rea$, we define the differential operator $\nabla H(x):=\left(\frac{\displaystyle \partial H}{\displaystyle \partial x}\right)^\top $ and for mappings $W:  \rea^n \times \rea^m\to \rea$ we define  $\nabla_y W(x,y):=\left(\frac{\displaystyle \partial W}{\displaystyle \partial y}\right)^\top $. 
 
\section{Problem Formulation}
\label{sec2}
%
In this section we formulate the algebraic state observation problem addressed in the paper. First, the mathematical model of the system that we consider is presented. Then, the algebraic observation scenario adopted in the paper is presented. 
\subsection{Mathematical model}
\label{subsec21}
\begin{figure}
	\centering
	\includegraphics{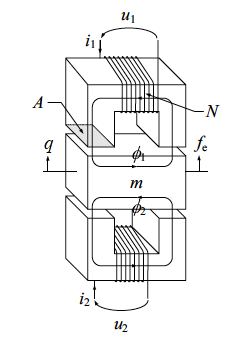}
	\caption{Opposed electromagnets: a single axis AMB supporting a mass $m$.}
	\label{fig1}
\end{figure}

%
In this paper we consider the single axis active magnetic bearing (AMB) depicted in Fig. \ref{fig1} \cite{MASbook}. The electrical port variables are $(v,i)$, where $v(t) \in \rea^2$ are the voltages and $i(t) \in \rea^2$ the currents. It is assumed that there are external voltage sources through which electrical energy is supplied to the magnetic elements, hence the voltages are the control signals. The mechanical port variables are  $(F,\dot q)$, where $F(t)\in \rea$ are the mechanical forces of electrical origin and  $\dot q(t) \in \rea$ is the translational velocity of the movable mass.  

The magnetic energy stored in the inductances is defined by the function $H_E(\phi,q)$, where $\phi(t) \in \rea^2$ is the vector of flux linkages and is given by
$$
H_E(\phi,q)  = {1 \over c_1}[ (c_2-q)\phi_1^2 +  (c_2+q)\phi_2^2],
$$
where $c_1,c_2$ are positive constants defined in \cite{MASbook}. The mechanical energy stored by the movable inertia is given by the function
$$
H_M(\mathfrak p)  =  {1 \over 2m} {\mathfrak p}^2,
$$
where ${\mathfrak p}(t) \in \rea$ is the momenta of the mass $m$.  

The constitutive relations of the elements are
\begequarrs
i & = & \nabla_\phi H_E(\phi,q)\\
F & = & - \nabla_q H_E(\phi,q)\\
\dot q & = &  \nabla H_M({\mathfrak p}),
\endequarrs 
where the minus sign in the force of electrical origin of the second equation reflects Newton's third law. The equations of motion of the system can be described in port-Hamiltonian (pH) form \cite{VANbook} as\footnote{To simplify the presentation, we disregard the presence of an external disturbance force in the mechanical dynamics---denoted $f_e$ in  \cite[eq. (15.7a)]{MASbook}---which can be easily incorporated in the observer design, even if its not measurable.}  
\begalis{
	\begmat{\dot \phi_1 \\ \dot \phi_2 \\ \dot q \\ \dot {\mathfrak p}} & =\begmat{-{R \over N} {\bf I}_2 & {\bf 0}_{2 \times 1} & {\bf 0}_{2 \times 1} \\ {\bf 0}_{1 \times 2} & 0 & 1 \\ {\bf 0}_{1 \times 2} & -1 & 0} \nabla H(\phi,q,{\mathfrak p}) + \begmat{{1 \over N}v\\ 0 \\ 0}\!,
}
where we have defined the systems total energy function
$$
H(\phi,q,\mathfrak p):= H_E(\phi,q)+H_M({\mathfrak p}),
$$
we assumed the presence of resistive elements in series with the inductances, with $R>0$ the value of the resistor, and $N$  the number of turns of wire in the coil.

To simplify the derivation of the observer we find convenient to express the AMB dynamics in the following state-space form \cite[eqs. (15.4)]{MASbook}:
\begsubequ
\label{stamod}
\begali{
	\label{stamod1}
	\dot \phi_1&=-{R \over N}y_1{+}{1 \over N}u_1\\
	\label{stamod2}
	\dot \phi_2&=-{R \over N}y_2{+}{1 \over N}u_2\\
	\label{stamod3}
	\dot q&={1 \over m}{\mathfrak p} \\
	\label{stamod4}
	\dot {\mathfrak p}&={N \over c_1}(\phi_1^2-\phi_2^2)\\
	\label{stamod5}
	y_1&= {2 \over c_1}(c_2-q)\phi_1 \\
	\label{stamod6}
	y_2&= {2 \over c_1}(c_2+q)\phi_2,
}
\endsubequ
where we defined the input and output signals  $u(t):=v(t) \in \rea^2,\;y(t):=i(t) \in \rea^2.$ We underscore the fact that, as seen in \eqref{stamod1} and \eqref{stamod2}. the {\em derivative} of the flux state components are available for {\em measurement}. As explained in \cite{BOBetal_tac}, this is an indispensable requirement for the design of an ASLO, which is exploited via the application of the SL. 
\subsection{Formulation of the algebraic state observation problem}
\label{subsec22}
%
Consider the AMB system \eqref{stamod}, where the only signals available for measurement are the currents $y$ and the voltage sources $u$ and the parameters $c_1,c_2,R$ and $N$ are known. In this paper we take a radically different approach to the state observation problem---our objective being to obtain an {\em algebraic} relation between the {\em unmeasurable} state vector and {\em filtered} versions of the systems inputs and outputs, with this identity holding true {\em for all $t \geq 0$}.  That is, we want to find a {\em mapping} $\calo: \rea^{6} \to \rea^{4}$ such that the following identity holds true\footnote{The claim that the identity \eqref{staide} {\em holds true} for all $t \geq 0$ should be understood taking into account that the action of the filters {\em is not} instantaneous. Since the filters time constants can be chosen arbitrarily small, with an obvious abuse of notation, we {\em neglect} its transient.}
\begequ
\label{staide}
x(t)=\calo[u(t),y(t),\calf[u](t),\calf[y](t)],\;\forall t \geq 0,
\endequ
where $x:=\col(\phi_1,\phi_2,q,\mathfrak p)$ and $\calf(p) \in \rea(p)$ is a---{\em designer chosen}---linear time-invariant (LTI) filter. Notice that, in contrast to the standard definition of an observer, we want the identity \eqref{staide} to hold true {\em for all $t \geq 0$}, and not {\em asymptotically} as $t \to \infty$.

\section{Main Result}
\label{sec3}
%
In this section we present the ASLO that solves the problem posed in Subsection \ref{subsec22}. Also, as indicated in the Introduction, because of singularity problems in the explicit calculation of the ASLO, we also present an A-ASLO---that avoids these problems and is directly derived from the ASLO.

\begin{proposition} \em
\label{pro1}
Consider the dynamics of the AMB system given in \eqref{stamod}. Define the LTI filter
\begequ
\label{fil}
\calf(p)={\lambda \over p + \lambda},\;\lambda>0,
\endequ
the following {\em measurable} signals
\begsubequ
\label{z}
\begali{
	z_1&:=-{R \over N}y_1{+}{1 \over N}u_1\\
	z_2&:=-{R \over N}y_2{+}{1 \over N}u_2,
}
\endsubequ
and the dynamic extension
\begsubequ
\label{dyn}
\begali{
	\label{dyn1}
	w_3&=\calf[y_2]+{4c_2 \over c_1 \lambda}\calf[z_2]-y_2\\
	\label{dyn2}
	w_4&=\calf[y_1]-{4c_2 \over c_1 \lambda}\calf[z_1]-y_1\\
	\label{dyn3}
	w_5&=\calf[w_3]\\
	\label{dyn4}
	w_6&=\calf[w_4]\\
	\label{dyn5}
	w_7&= {1 \over \lambda}\calf[z_2\calf[y_1]]+{1 \over \lambda}\calf[z_1\calf[y_2]]+\frac{4c_2}{c_1}{1 \over \lambda}\calf[z_2{1 \over \lambda}\calf[z_1]] \\
	&+\frac{4c_2}{c_1}{1 \over \lambda}\calf[z_1{1 \over \lambda}\calf[z_2]] \\
	\label{dyn6}
	w_8&=\calf[w_7]+{1 \over \lambda}\calf[z_1\calf[w_3]]+{1 \over \lambda}\calf[z_2\calf[w_4]].
}
\endsubequ

\noindent {\bf C1} Define the scalar signal 
$$
\Delta:=w_3w_6-w_4w_5.
$$
Assume $\Delta(t) \neq 0$. Then, an ASLO of the system is given by
\begali{
	\label{aslophi}
	\phi(t)= {1 \over \Delta(t)}\begmat{ w_6(t) w_7(t)-w_4(t) w_8(t) \\ w_3(t)  w_8(t) -w_5(t) w_7(t) },\;\forall t \geq 0.
}
\noindent {\bf C2}  Given the system \eqref{stamod} and the algebric state observer \eqref{aslophi} we define an A-ASLO for the flux coordinates as:
\begsubequ
\label{obsphi}
\begali{
	\label{obsphi1}
	\dot {\hat {\phi}}_1 &=z_1-\gamma_1\Delta^2 \hat \phi_1+ \gamma_1 \Delta (w_6w_7-w_4w_8)\\
	\label{obsphi2}
	\dot {\hat {\phi}}_2&=z_2-\gamma_2\Delta^2 \hat \phi_2+ \gamma_2 \Delta (w_3w_8-w_5w_7),
}
\endsubequ
where $\gamma_1,\gamma_2>0$. The following equivalence holds true
$$
\Delta(t) \notin \call_2\;\iff\;\liminf |\hat \phi(t)-\phi(t)|=0.
$$

\noindent {\bf C3} Moreover, an A-ASLO for the position and velocity coordinates is given by
\begsubequ
\label{obsqp}
\begali{
	\label{obsq}
	\dot{\hat{q}}&=\frac{1}{m}\hat {\mathfrak p} -l_1\hat q+l_1\frac{z_3}{z_4} \\
	\label{obsp}
	\dot{\hat {\mathfrak p}}&=\frac{N}{c_1}(\hat \phi_1^2-\hat \phi_2^2)-l_2\hat q+l_1\frac{z_3}{z_4},
}
\endsubequ
with $l_1,l_2>0$, where we defined the measurable signals
\begali{
	\nonumber
	z_3 &:= \frac{2c_2}{c_1}(\hat \phi_1^2-\hat \phi_2^2)+y_2\hat \phi_2 - y_1 \hat \phi_1\\
	\label{z3z4}
	z_4 &:= \frac{2}{c_1}(\hat \phi_1^2 + \hat \phi_2^2),
}
and $\hat \phi$ is given by \eqref{obsphi}. Assume the {\em generic assumption} $|\hat \phi(t)|\neq 0,\forall t \geq 0$, is satisfied. Then, the following implication is true
$$
\Delta(t) \notin \call_2  \implies \begmat{&\liminf |\hat q(t)-q(t)|  =0 \\ & \\
	& \liminf |\hat {\mathfrak p}(t)-{\mathfrak p}(t)| =0}.
$$
\qed
\endpro

Before closing this section we make the observation that, the new A-ASLO requires the division by $|\hat \phi(t)|$, which is consistent with the operation of the physical system. This should be contrasted with the ASLO that imposes the rather artificial constraint of $\Delta(t) \neq 0$. 

\section{Simulation Results}
\label{sec4}
%
The 2-DOF AMB system \eqref{stamod} was simulated with the following parameters of the system and initial conditions:
\begin{align*}
	R = 1.6, N = 321, c_1=c_2=293.5, m=2.5, \\
	\phi_1(0) = 0.002, \phi_2(0)=0.002, q(0) = 0.05, \mathfrak{p}(0)=0.
\end{align*}

The 2-DOF AMB system \eqref{stamod} with the A-ASLO \eqref{obsphi}, \eqref{obsqp} was simulated with different input signals.

The parameters of the position and velocity observer \eqref{obsqp} were parameterized as:
$$
l_1=2w_0, l_2 = mw_0^2 \text{ with some constant } w_0>0.
$$

Figures \ref{fig:phi1}, \ref{fig:phi2} shows the transient of the flux observation errors for the constant input signal $u_1=u_2=1$ with different values of the gains $\gamma_1$ and $\gamma_2$.

Figures \ref{fig:q}, \ref{fig:p} depicts the transient of the position and velocity observation errors for the constant input signal $u_1=u_2=1$ with different values of the parameter $w_0$. 

\begin{figure}
	\centering
	\includegraphics [width=1\linewidth]{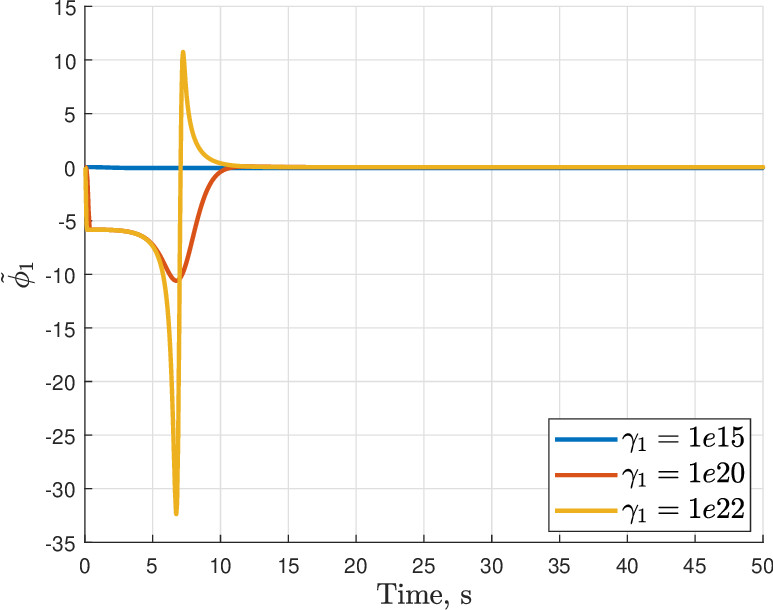}
	\caption{Transient of the estimation error $\tilde{\phi}_1$ with control signals $u_1=u_2=1$ }
	\label{fig:phi1}
\end{figure}

\begin{figure}
	\centering
	\includegraphics [width=1\linewidth]{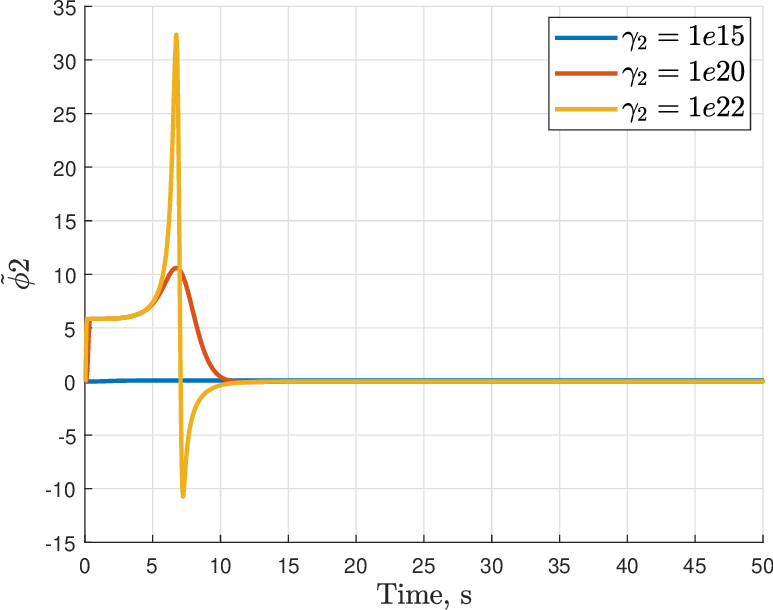}
	\caption{Transient of the estimation error  $\tilde{\phi}_2$ with control signals $u_1=u_2=1$ }
	\label{fig:phi2}
\end{figure}

\begin{figure}
	\centering
	\includegraphics [width=1\linewidth]{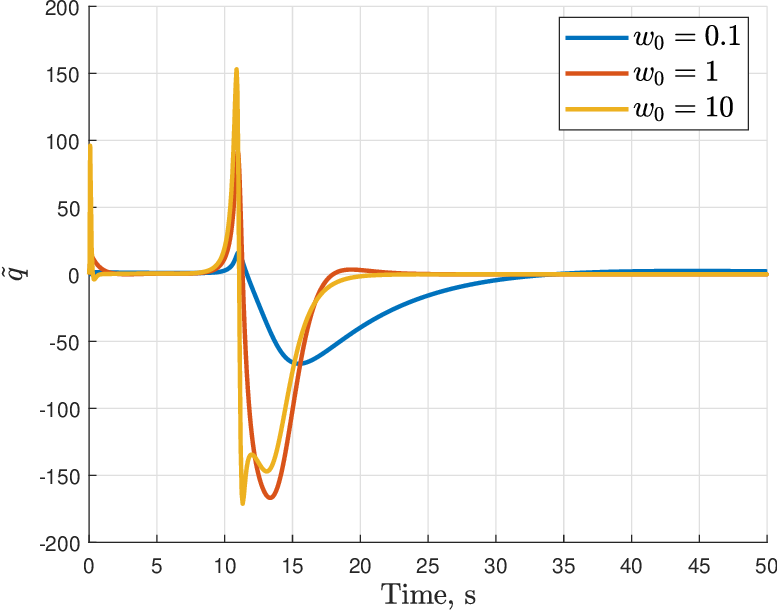}
	\caption{Transient of the estimation error $\tilde{q}$ with control signals $u_1=u_2=1$ }
	\label{fig:q}
\end{figure}

\begin{figure}
	\centering
	\includegraphics [width=1\linewidth]{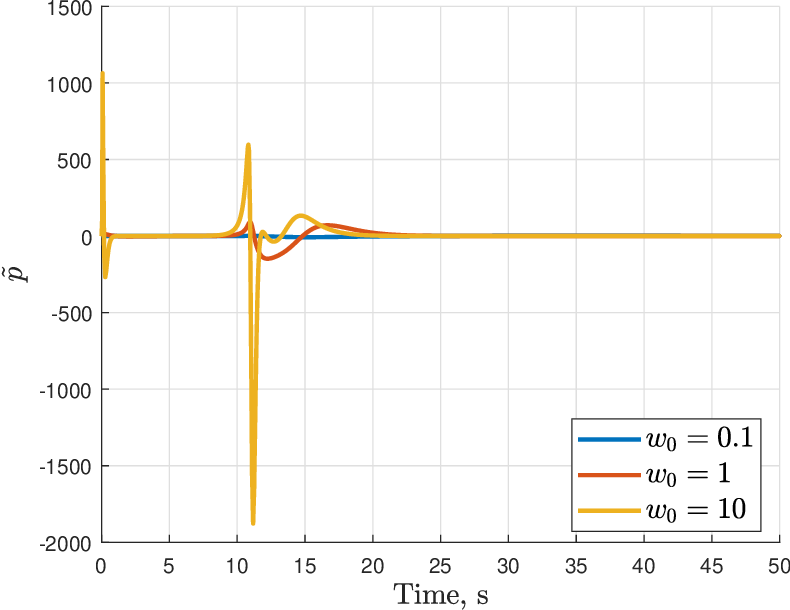}
	\caption{Transient of the estimation errors $\tilde{\mathfrak{p}}$ with control signals $u_1=u_2=1$ }
	\label{fig:p}
\end{figure}

Figures \ref{fig:phi12}, \ref{fig:phi22} demonstrates the transient of the flux observation errors for the input signals $u_1=50\sin(5t)$ and $u_2=30\sin(7t)$ with different values of the gains $\gamma_1$ and $\gamma_2$.

For these three cases the parameter of the LTI filter was  $\lambda = 1$.

Figures \ref{fig:q2},\ref{fig:p2} shows the transient of the position and velocity observation errors for the input signals $u_1=50\sin(5t)$ and $u_2=30\sin(7t)$ with different values of the parameter $w_0$. 
The parameter of the LTI filter for this case was $	\lambda = 3$.

\begin{figure}
	\centering
	\includegraphics [width=1\linewidth]{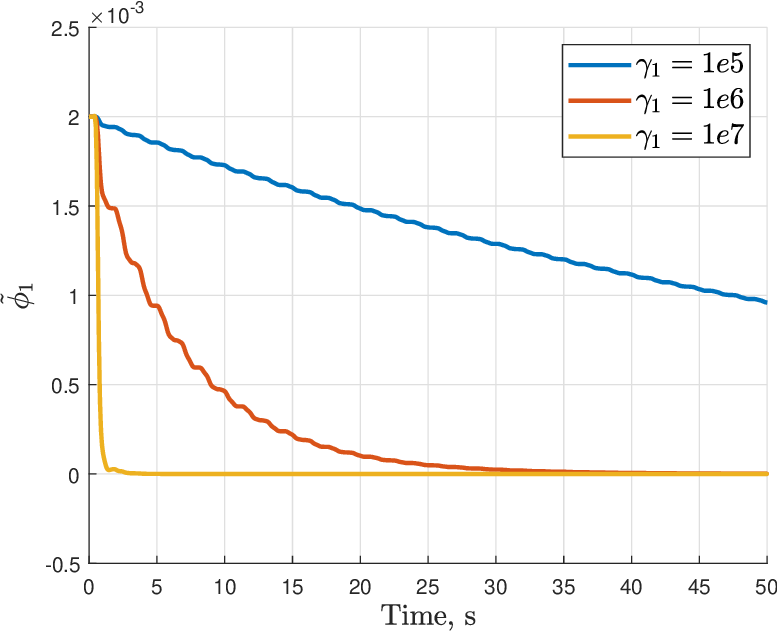}
	\caption{Transient of the estimation error $\tilde{\phi}_1$ with control signals $u_1=50\sin(5t)$ and $u_2=30\sin(7t)$ }
	\label{fig:phi12}
\end{figure}

\begin{figure}
	\centering

	\includegraphics [width=1\linewidth]{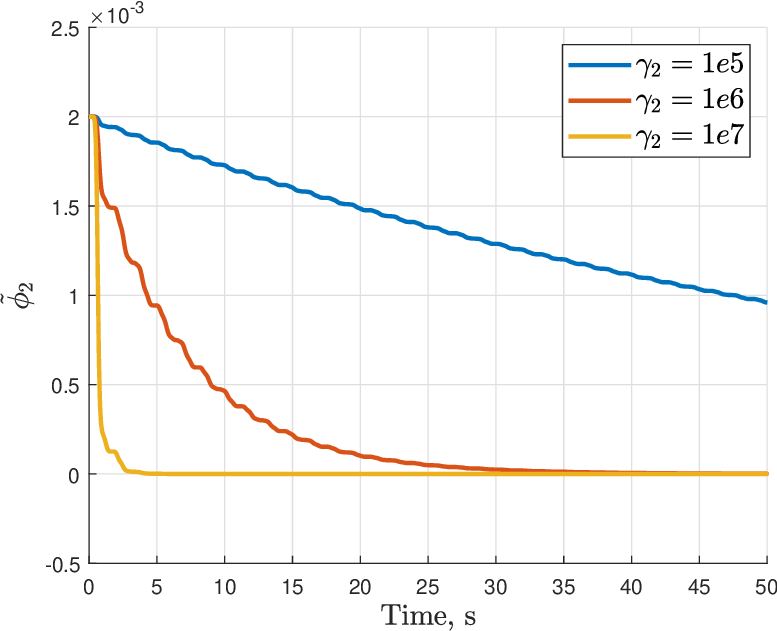}
	\caption{Transient of the estimation error $\tilde{\phi}_2$ with control signals $u_1=50\sin(5t)$ and $u_2=30\sin(7t)$ }
	\label{fig:phi22}
\end{figure}

\begin{figure}
	\centering
	\includegraphics [width=1\linewidth]{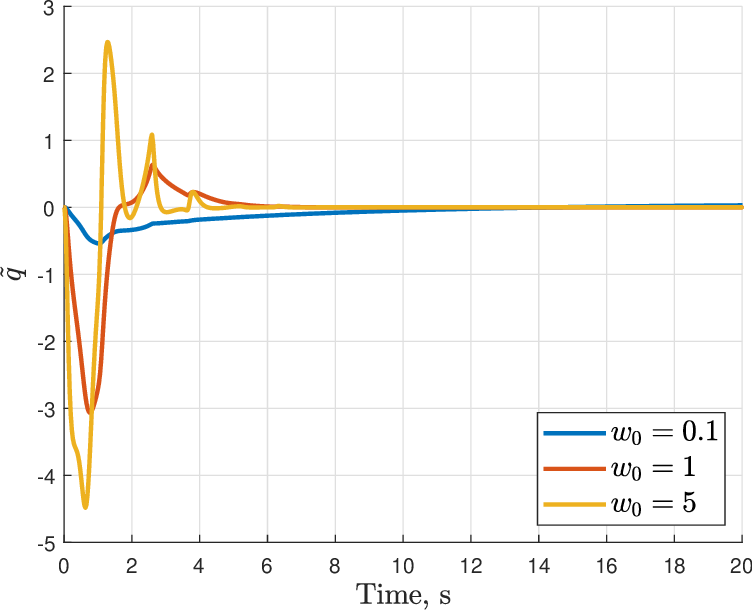}
	\caption{Transient of the estimation errors $\tilde{q}$ with control signals $u_1=50\sin(5t)$ and $u_2=30\sin(7t)$}
	\label{fig:q2}
\end{figure}

\begin{figure}
	\centering
	\includegraphics [width=1\linewidth]{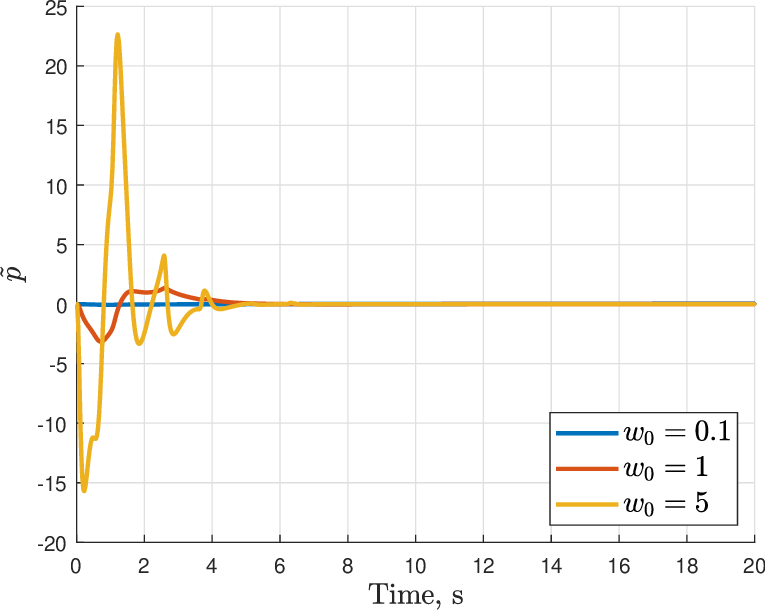}
	\caption{Transient of the estimation errors $\tilde{\mathfrak{p}}$ with control signals $u_1=50\sin(5t)$ and $u_2=30\sin(7t)$}
	\label{fig:p2}
\end{figure}

Comparing the figures with the input $u=(1,1)$ and the ones with $u(t)=(50\sin(5t),30\sin(7t))$ we see the beneficial effect of exciting the system.

\section{Conclusions and Future Work}
\label{sec5}
%
We have presented, for the first time, an {\em algebraic} observer for the estimation of the state of an AMB. Unfortunately, due to the poor ``excitation" conditions, the performance of the ASLO is below par. Therefore, to overcome this problem we have added an A-ASLO that is built directly from the original ASLO, an avoids the singularity problems.  It is interesting that this problem was not observed in the various physical examples reported in \cite{BOBetal_tac,BOBetal_jpc}, which include PMSM, induction motors, Ball-and-Beam and robotic leg systems and reaction systems. \\

Among the future research problems that we will address are: \begite
\item An adaptive version of the ASLO estimating the parameters $R$ and $f_e$
\item Carry out some experimental validation of the result.
\item Try to improve the performance of the ASLO, probably invoking the alternative variation of the SL reported in \cite{CHOetal}. 
\endite


\bmsubsection*{Conflicts of Interest}

The authors declare no conflicts of interest.

\bibliography{mbs}

\appendix

\bmsection{Swapping Lemma \cite[Lemma 3.6.5]{SASBODbook}}
\lab{appa}
%
Consider the LTI filter \eqref{fil}. Given the signals $w:\rea_+ \to \rea^{n \times m}$ and $v:\rea_+ \to \rea^m$ and the stable LTI filter \eqref{fil}. The following identity holds true
\begin{equation}
	\label{swalem}
	\calf[w v]= \calf[w] v-  \calf\Big[ {1 \over \lambda}\calf [w]\dot v\Big].
\end{equation}
\bmsection{Proof of Proposition \ref{pro1}}
\lab{appb}
%
\bmsubsection {Proof of claim C1} The main objective of the proof is to establish and algebraic relation of the form
\begequ
\lab{algsys}
\cala(t) \phi(t) = \calb(t),
\endequ
where $\cala(t) \in \rea^{2 \times 2}$ and $\calb(t) \in \rea^2$ are measurable matrices, with $\cala(t)$ invertible. Interestingly, this is possible via succesive applications of the well-known {\em Swapping Lemma} (SL) \cite[Lemma 3.6.5]{SASBODbook}, recalled for ease of reference in Appendix \ref{appa}.\footnote{To simplify the notation, throughout the proof we introduce signals denoted $w_i$, which are {\em measurable} }

Consider equations \eqref{stamod5} and \eqref{stamod6}. Multiplying the first one by $\phi_2$ and the second one by $\phi_1$, and adding them up we get 
\begin{equation}
	y_2 \phi_1 + y_1 \phi_2 =\frac{4c_2}{c_1}\phi_1 \phi_2
	\label{y1}
\end{equation}
Applying the SL to each of the terms of \eqref{y1} we obtain
\begalis{
	\calf[y_1 \phi_2] &=\calf[\phi_2 y_1]=\phi_2\calf[y_1]-{1 \over \lambda}\calf[z_2\calf[y_1]]\\
	\calf[y_2 \phi_1] &=\calf[\phi_1 y_2]=\phi_1\calf[y_2]-{1 \over \lambda}\calf[z_1\calf[y_2]]\\
	\calf[1 \times (\phi_1 \phi_2)]&=\phi_1 \phi_2\calf[1]-{1 \over \lambda}\calf[(\dot \phi_1 \phi_2+\phi_1 \dot \phi_2)\calf[1]] \\
	&=\phi_1 \phi_2 - {1 \over \lambda}\calf[z_1\phi_2+\phi_1 z_2] \\
	& =\phi_1\phi_2-(\phi_2{1 \over \lambda}\calf[z_1]-{1 \over \lambda}\calf[z_2{1 \over \lambda}\calf[z_1]]) \\
	&-(\phi_1{1 \over \lambda}\calf[z_2]-{1 \over \lambda}\calf[z_1{1 \over \lambda}\calf[z_2]]) \\
	&= \phi_1 \phi_2 - \phi_2{1 \over \lambda}\calf[z_1]+{1 \over \lambda}\calf[z_2{1 \over \lambda}\calf[z_1]] \\
	&-\phi_1{1 \over \lambda}\calf[z_2]+{1 \over \lambda}\calf[z_1{1 \over \lambda}\calf[z_2]].
}
Thus after applying filter $\calf$ to \eqref{y1}, and using the equations above, we obtain
\begequ
\lab{ff}
\begin{aligned}
	&\phi_2\Big(\calf[y_1]+\frac{4c_2}{c_1}{1 \over \lambda}\calf[z_1]\Big)+\phi_1\Big(\calf[y_2]+\frac{4c_2}{c_1}{1 \over \lambda}\calf[z_2]\Big)-\frac{4c_2}{c_1}\phi_1 \phi_2 \\
	&={1 \over \lambda}\calf[z_2\calf[y_1]]+{1 \over \lambda}\calf[z_1\calf[y_2]]+\frac{4c_2}{c_1}{1 \over \lambda}\calf[z_2{1 \over \lambda}\calf[z_1]] \\
	&+\frac{4c_2}{c_1}{1 \over \lambda}\calf[z_1{1 \over \lambda}\calf[z_2]]
\end{aligned}
\endequ

To simplify the notation let us define variables:
\begalis{
	w_7 &:= {1 \over \lambda}\calf[z_2\calf[y_1]]+{1 \over \lambda}\calf[z_1\calf[y_2]]+\frac{4c_2}{c_1}{1 \over \lambda}\calf[z_2{1 \over \lambda}\calf[z_1]] \\
	&+\frac{4c_2}{c_1}{1 \over \lambda}\calf[z_1{1 \over \lambda}\calf[z_2]] \\
	w_1 &:= \calf[y_2]+\frac{4c_2}{c_1}{1 \over \lambda}\calf[z_2] \\
	w_2 &:= \calf[y_1]+\frac{4c_2}{c_1}{1 \over \lambda}\calf[z_1].
}
Notice that these three signals, $w_1, w_2$ and $w_7$, are {\em measurable}.

Using the definitions above in \eqref{ff} we obtain
\begin{equation}
	\label{y3}
	\phi_2w_2+\phi_1w_1-\frac{4c_2}{c_1}\phi_1\phi_2=w_7
\end{equation}
Now, we can write the system of two equations \eqref{y1} and \eqref{y3} as
$$
\begin{cases}
	y_1\phi_2+y_2\phi_1-\frac{4c_2}{c_1}\phi_1\phi_2= 0\\
	\phi_1w_1+\phi_2w_2-\frac{4c_2}{c_1}\phi_1\phi_2=w_7 \\
\end{cases}
$$
After subtraction we obtain: 
\begequ
\lab{w7}
w_7=w_3\phi_1+w_4\phi_2,
\endequ
where we defined $w_3:= w_1 - y_2$ and $w_4:= w_2-y_1$, which are, clearly, {\em measurable}.

Now we can apply filter $\calf$ to the equation above:
$$
\calf[w_7]=\calf[\phi_1w_3]+\calf[\phi_2w_4]
$$
Applying the SL we can rewrite this equation as
$$
\calf[w_7]=\phi_1\calf[w_3]-{1 \over \lambda}\calf[z_1\calf[w_3]]+\phi_2\calf[w_4]-{1 \over \lambda}\calf[z_2\calf[w_4]]
$$
The resulting equation can be represented in the following form:
\begin{equation}
	\lab{w8}
	w_8=w_5\phi_1+w_6\phi_2
\end{equation}
with the definitions:
\begalis{
	w_8&:=\calf[w_7]+{1 \over \lambda}\calf[z_1\calf[w_3]]+{1 \over \lambda}\calf[z_2\calf[w_4]]\\
	w_5 &:=\calf[w_3]\\
	w_6 &:=\calf[w_4]
}

The equations \eqref{w7} and \eqref{w8} constitute the algebraic system \eqref{algsys} we are looking for, with the definitions
\begalis{
	\cala &:= \begmat{w_3 & w_4 \\ w_5 & w_6},\;\calb :=\begmat{ w_7 \\ w_8}.
}
Under the {\em generic assumption} that $w_3(t)w_6(t)-w_4(t)w_5(t)\neq 0$, we can solve this system to get
\begin{equation}
	\label{y56}
	\begin{cases}
		w_6w_7-w_4w_8=\Delta\phi_1\\
		-w_5w_7+w_3w_8=\Delta\phi_2\\
	\end{cases}
\end{equation}
where we defined the matrix determinant $\Delta:=w_3w_6-w_4w_5$. This completes the proof of claim {\bf C1}.\\

\bmsubsection  {Proof of claim C2} Define the observation error $\tilde \phi:=\hat \phi - \phi$. Using the fact that $(z_1,z_2)$---defined in \eqref{z}---is equal to $\dot \phi$ and taking into account \eqref{y56}, we conclude that the error equation for the flux A-ASLO \eqref{obsphi} satisfies
$$
\begin{cases}
	\dot {\tilde \phi}_1=-\gamma_1\Delta^2 \tilde \phi_1\\
	\dot {\tilde \phi}_2=-\gamma_2\Delta^2 \tilde \phi_2,
\end{cases}
$$
which completes the proof.\\

\bmsubsection {Proof of claim C3} First, we make the observation that from \eqref{stamod5} and \eqref{stamod6} we get the relations
\begalis{
	q \phi_1&=-{c_1 \over 2}y_1+c_2 \phi_1\\
	q \phi_2&=-{c_1 \over 2}y_2-c_2 \phi_2.
}
After some algebraic operations with these equations we obtain the key relation
\begequ
\lab{keyrel}
{2 \over c_1}q={1 \over |\phi|^2}\Big[{2 c_2 \over c_1}(\phi^2_1 -\phi^2_2)+(y_2\phi_2-y_1\phi_1)\Big].
\endequ
We notice now that, if we {\em temporarily} evaluate $z_3$ and $z_4$ with $\hat \phi=\phi$, and replace these expressions in \eqref{keyrel} we obtain
\begequ
\lab{z3ovez4}
{z_3 \over z_4}=q.
\endequ

Define the estimation error for the mechanical coordinates of the A-ASLO \eqref{obsqp}:
$$
\tilde q :=\hat q -q,\;\tilde{\mathfrak{p}} := \hat{\mathfrak{p}}-\mathfrak{p}.
$$
Now, recalling that the momenta is defined as $\mathfrak{p}=m\dot q$, naturally denoting $\hat {\mathfrak{p}}=m\hat{\dot q}$, and using \eqref{z3ovez4} we can write \eqref{obsq} as
$$
\dot {\tilde q} =\frac{1}{m}\tilde {\mathfrak p} -l_1 \tilde  q.
$$

Regarding the momenta estimator equation, we recall from \eqref{stamod4} that
$$
\dot {\mathfrak p}={N \over c_1}(\phi_1^2-\phi_2^2).
$$
Hence, evaluating (again temporarily) \eqref{obsp} with $\hat \phi=\phi$, we obtain the error equation for the momenta as
$$
\dot {\tilde {\mathfrak {p}}}=-l_2\tilde q. 
$$
Combining the two error equations we finally get
$$
\begmat{\dot{\tilde q} \\ \dot{\tilde{\mathfrak p}}}=\begmat{ -l_1 &\frac{1}{m} \\ -l_2 & 0 }\begmat{ \tilde q \\ \tilde{\mathfrak p}}
$$
which is an asymptotically stable system. 

To complete the proof we repeat the calculations above denoting
$$
\hat \phi=\phi + \tilde \phi,
$$
that, under the standing assumptions verifies $\tilde \phi \to 0$, and dragging the decaying term throughout. These operations are standard in certainty-equivalent based designs and are omitted for brevity.


\end{document}

%% file: aux_format.tex
\def\begcen{\begin{center}}
\def\endcen{\end{center}}

\newcommand{\col}{ \mbox{col} }

\def\calo{{\cal O}}

\def\calb{{\cal B}}

\def\call{{\cal L}}
\def\calf{{\cal F}}
\def\cala{{\cal A}}

\def\liminf{\lim_{t \to \infty}}

\def\L2{{\cal L}_2}
\def\L2e{{\cal L}_{2e}}

\def\rea{\mathbb{R}}

\def\begmat#1{\begin{bmatrix}#1\end{bmatrix}}
\def\begali#1{\begin{align}{#1}\end{align}}
\def\begalis#1{\begin{align*}{#1}\end{align*}}

\def\begequarr{\begin{eqnarray}}
\def\endequarr{\end{eqnarray}}
\def\begequarrs{\begin{eqnarray*}}
\def\endequarrs{\end{eqnarray*}}
\def\begarr{\begin{array}}
\def\endarr{\end{array}}
\def\begequ{\begin{equation}}
\def\endequ{\end{equation}}
\def\lab{\label}
\def\begdes{\begin{description}}
\def\enddes{\end{description}}
\def\begenu{\begin{enumerate}}
\def\begite{\begin{itemize}}
\def\endite{\end{itemize}}
\def\endenu{\end{enumerate}}

\def\lef[{\left[\begin{array}}
\def\rig]{\end{array}\right]}
\def\qed{\hfill$\Box \Box \Box$}
\def\begcen{\begin{center}}
\def\endcen{\end{center}}
\def\begrem{\begin{remark}\rm}
\def\endrem{\end{remark}}
\def\begassum{\begin{assumption}}
\def\endassum{\end{assumption}}
\def\begassums{\begin{assumption*}}
\def\endassums{\end{assumption*}}
\def\begassu{\begin{ass}}
\def\endassu{\end{ass}}
\def\beglem{\begin{lemma}}
\def\endlem{\end{lemma}}
\def\begcor{\begin{corollary}}
\def\endcor{\end{corollary}}
\def\begfac{\begin{fact}}
\def\endfac{\end{fact}}

\def\liminf{\lim_{t \to \infty}}

\def\L2e{{\cal L}_{2e}}

\def\rea{\mathbb{R}}

\def\col{\mbox{col}}

\def\begsubequ{\begin{subequations}}
	\def\endsubequ{\end{subequations}}
\def\begpro{\begin{proposition}}
	\def\endpro{\end{proposition}}
\def\beglem{\begin{lemma}}
	\def\endlem{\end{lemma}}
\def\begass{\begin{assumption}}
	\def\endass{\end{assumption}}
\def\begcor{\begin{corollary}}
	\def\endcor{\end{corollary}}
\def\begproo{\begin{proof}}
	\def\endproo{\end{proof}}

\usepackage{xcolor}